\documentclass[fleqn, twoside]{article}
\usepackage{espcrc2}

\usepackage{graphicx}
\usepackage[figuresright]{rotating}

\newcommand{\AmS}{{\protect\the\textfont2
  A\kern-.1667em\lower.5ex\hbox{M}\kern-.125emS}}

\def\Dslash{\not{\hbox{\kern-4pt $D$}}} 

\newcommand {\beq}{\begin{equation}}  
\newcommand {\eeq} {\end{equation}}

\newcommand {\bea} {\begin{eqnarray}} 
\newcommand {\eea} {\end{eqnarray}}

\newcommand {\ber} {\begin{eqnarray*}} 
\newcommand {\eer} {\end{eqnarray*}} 

\newcommand {\state} [1] {\mid \! \! {#1} \rangle} 

\title{THE ADJOINT VACUUM SECTOR OF MASSLESS  QCD2}

\author{A. Abrashkin\address[tau]{Physics and Astronomy, Tel Aviv University},
Y. Frishman\address{Particle Physics, Weizmann Institute}
\thanks{Gave talk at the Workshop}, 
J. Sonnenschein\addressmark[tau]}

\begin{document}

\begin{abstract}
\noindent
We consider a ``one current'' state, obtained by application of a   
color current on the ``adjoint'' vacuum, in $QCD_2$, with   
quarks in fundamental representation. 

\noindent
The quarks  are taken to be  massless.
The theory  on  the light-front can be ``currentized'', namely   
formulated in terms of currents only.  

\noindent
The adjoint vacuum is obtained by applying a current derivative,  
at zero momentum, on the singlet vacuum.  

\noindent
In general the ``one current'' states are not 
eigenstates of $M^2=2P^+P^-$, except in the large $N_f$ limit. 

\noindent
Problems with infra-red regularizations are pointed out.  

\noindent
Connection to fermionic structure is made.

\noindent  
This talk is based on the paper Ref \cite{abfs}, by the same authors.

\vspace{1pc}
\end{abstract}

\maketitle

\section{Introduction}

QCD2  in terms of only  colored currents (``currentization''), 
turns out to be very natural once the system is quantized on the light-front. 

Both the momentum and the  
Hamiltonian, and hence also $M^2$, are expressed in terms of the 
light-cone colored currents. 

In fact only  
the left (or right) currents are needed.

The currentization was shown to hold for multiflavor fundamental and 
adjoint quarks. 

In fact, it can be applied whenever 
the free fermions energy momentum tensor can be written in term of a 
Sugawara form.  

The light-cone momentum and Hamiltonian of  
any CFT that posseses an affine Lie algebra and is coupled to non Abelian 
gauge 
fields associated with the  
same algebra, can also be described in terms of holomorphic currents.

The Fock space of physical states is obtained by applying current 
creation operators on the 
vacuum.   

The lowest  
physical states constructed by applying  current creation operators on the 
{\bf singlet vacuum}, are those 
built from two currents. 

A 't Hooft like equation for the wave function of these  
two currents states was obtained, and solved   
for the lowest massive state, in Ref \cite{arfs}.

Excellent agreement with the DLCQ results, Refs \cite{ghk}, \cite{ap}, 
\cite{t}. 

Turns  out that one can also construct states by using only  one current 
creation operators,  
applied on the {\bf adjoint vacuum}.  

The latter is obtained 
by acting on  the singlet vacuum with fermionic zero modes. 

In the case of adjoint fermions, by a single  
adjoint zero mode, and for fundamental fermions by quark anti-quark zero modes.
  
In a scheme where only currents are being used one should be able to express 
the adjoint vacuum also in terms 
of currents, which we do, as mentioned before, for the case of fundamental 
quarks. 
 
The outcome of our analysis is that in the large $N_f$ limit the state of one 
current 
applied on the adjoint vacuum, 
is indeed an eigenstate of the mass operator with a mass of 
$\sqrt{e^2 N_f\over \pi}$.

In the large $N_c$ limit this is not the case.
Acting with $M^2$ 
on this state, the dominant daughter state is a two current state. 

We further analyze the fermionic structure of these states,
especially in the large $N_c$ limit, to connect to the 't Hooft analysis.
 
En route to these results we are faced with technical obstacles, due to   
infra red (IR) divergences. 

In certain parts of the computations 
we were able to regularize the IR divergences, but in others we left it as an 
open problem. 

Assuming 
that a regularization scheme can be found, we have fully determined the 
normalizations of the relevant states.   

\section{QCD2 Currentization} 

The action for multi-flavor massless  fermions, in the fundamental 
representation of $SU(N_c)$ 

\bea
\lefteqn{S=\int d^2x \ {\rm tr}\ (-{1\over 2e^2}F_{\mu\nu}^2+i\bar 
\Psi\Dslash\Psi)} \\ 
&& \Psi = \Psi ^i_a, i=1 \dots N_c, a=1 \dots N_f  \nonumber \\
&& D_\mu = \partial_\mu + i A_\mu  \nonumber
\eea

An alternative description is achieved by bosonizing the theory. 
The bosonization of  
multi-flavor massive ${\rm QCD}_2$ is complicated since one has to translate 
the fermions into bosonic variables which are group elements of  
$U(N_F\times N_c)$. See Ref \cite{fs1}.
 
For massless fermions one can use bosonization in the  
$SU(N_c)\times SU(N_f)\times U_B(1)$ scheme. 

This scheme is more convenient since  it  decouples  
the color and flavor degrees of freedom.  

For a Review see Ref \cite{fs2}.

The bosonized form of the action in this scheme is 

\bea
\lefteqn{S_{b} = N_f S_{WZW}(h) + N_c S_{WZW}(g)}  
\nonumber \\
&& + \int d^2x\ {1\over 2}\partial _\mu \phi \partial ^\mu
 \phi -\int d^2x\ {\rm tr}\ {1\over {2e^2}} F
 _{\mu \nu} F^{ \mu \nu} \nonumber
\\
&& -{N_f\over 2\pi} \int d^2 x \ {\rm tr}\ (ih^\dagger \partial_+ h A_-
+ih\partial_ - h^\dagger A_+ \nonumber \\
&& + A_+ h A_- h^\dagger - A_+ A_-)   
\eea

$h \in SU(N_c)$, $g\in SU(N_f)$, $\phi$ is the bosonic field for the baryon 
number, 
 and $S_{WZW}$ stands for the 
Wess-Zumino-Witten action

\bea
\lefteqn{S_{WZW}(g)={1\over{8\pi}}\int _\Sigma d^2x \ {\rm tr}\ (\partial _\mu
g\partial ^\mu g^{-1}) } \nonumber \\
 && + {1\over{12\pi}}\int _B d^3y \epsilon ^{ijk} \nonumber \\
 && {\rm tr}\ (g^{-1}\partial _i g) (g^{-1}\partial _j g)(g^{-1}\partial _k g)
\eea

 $B$ is a three dimensional volume whose boundary is the two dimensional 
surface $\Sigma$, which in our case is the 1+1 Minkowsky space.
 
The flavored sectors are indeed decoupled from the colored one. 
Moreover, the former ones are entirely massless.   
Since we are interested in the massive spectrum of the theory,  
we will set  aside the $g$ and $\phi$ fields and analyze only the colored 
field $h$. 

The residual interaction of the zero modes of the $g, h$ and $\phi$ fields, 
will  not be important to our discussion. 

Choose the light cone gauge $A _- = 0$, and quantize the system on the  
light-front $x_-=0$.
Integrating $A_+$ we get an effective non-local action

\bea
\lefteqn{S = N_f S_{WZW}(h) - {1\over 2} e^2 \int d^2x\ {\rm tr}\  
({1\over \partial _-} 
J^+)^2}   \\
&& J^+ = {i N_f\over 2\pi} h\partial _- h^{\dagger} \nonumber 
\eea  

The light-front momentum and energy
 take now simplified  forms. 

The momentum takes the Sugawara form

\bea\lefteqn
{P^+={1\over{N_c+N_f}}} \nonumber \\
&&  \int dx^- :J^a(x^-,x^+=0)J^a(x^-,x^+=0)
\eea

The energy

\bea
\lefteqn{P^-=-{e^2 \over {2\pi}}\int dx^-} \nonumber \\
&& :J^a(x^-,x^+=0){1\over{\partial _- 
^2}}J^a(x^-,x^+=0):  \\
&& J= \sqrt \pi J^+ \nonumber 
\eea

Evaluate mass
\beq
2P^+P^-\state{\psi} = M^2\state{\psi}. 
\eeq

\bigskip

Note:

1.
The equations that determine the spectrum are entirely expressed 
in terms of currents.   

2.
Both $P^+$ and $P^-$ depend only on $J^+$ and not on $J^-$.

3.
The only condition 
for   currentizing  a theory of fermions coupled to non Abelian gauge fields, is that 
$T^{++}$ of the free fermionic theory could be rewritten in terms of a Sugawara form. 

4.
In particular 
this can be done for adjoint fermions and 
for  fermions in the symmetric and antisymmetric ``two box'' 
representations.    
 
\bigskip

In momentum space
\beq
J(p^+)=\int {dx^- \over {\sqrt{2\pi}}} e^{-ip^+x^-} J(x^-,x^+=0)
\eeq

Normal ordering in the expression of $P^+$ and $P^-$ are naturally with 
respect 
to $p^+$, where $J(p^+ <0)$ denotes a creation operator. 
 To simplify the notation we 
write from now on $p$ instead of $p^+$. 
 
\beq
P^{+} = \frac{2}{N_{f}+N_{c}}\int_{0}^{\infty}dpJ^{b}(-p)J^{b}(p) 
\eeq

\beq
P^{-} = \frac{e^{2}}{2\pi}\int_{0}^{\infty}dp\Phi(p)J^{b}(-p)J^{b}(p)
\eeq

\beq
\Phi(p) = \frac{1}{2}\left(\frac{1}{(p+i\epsilon)^{2}}+ 
\frac{1}{(p-i\epsilon)^{2}}\right)
\eeq

$\Phi(p)$ is $(-\frac{\sqrt\pi}{2})$ times the Fourier transform of the 
'potential' $\left|x-y\right|$.

In computing the eigenvalues of 
$M^2=2P^+P^-$, 
we will need the algebra 
\bea
\lefteqn{[J^a(p),J^b(p')]=}  \nonumber \\
&& {1\over 2}N_f\ p\ \delta ^{ab} \delta (p+p')+ 
 if^{abc} J^c(p+p')
\eea

The vacuum $\state{0}$ obeys
\beq
\forall p>0,\ J(p)\state{0}=0 
\eeq

 Physical states are built by applying the current creation operators on the 
vacuum

\beq
\state{\psi} ={\rm tr}\ J(-p_1)\ldots J(-p_n)\state{0}
\eeq

Note that 
this basis is not orthogonal. 

\section{The One-Current State}

Define ``adjoint vacuum''
\beq
\left|0,\; ij \right\rangle =\lim_{\epsilon\rightarrow0}b_{\beta}^{\dagger i}(\epsilon) 
d_{j}^{\dagger\beta}(\epsilon)\left|0\right\rangle 
\eeq

$b_{\beta}^{\dagger i}$ and  $d_{j}^{\dagger\beta}$ are the creation operators 
of a quark and anti-quark respectively.

We can represent the action of the above adjoint zero mode on the 
vacuum by the derivative of a creation current taken at zero momentum. 

\bea
\lefteqn{J_{j}^{'i}(k)\left|0\right\rangle _{k=0^{-}} =
\sqrt{\frac{\pi}{2}} \frac{d}{dk}\int_{0}^{\infty}dp\int_{0}^{\infty}dq }
\nonumber \\ &&
\delta(k+p+q) b_{\beta}^{\dagger i}(p)d_{j}^{\dagger\beta}(q)\left|0\right\rangle _{k=0^{-}} 
\nonumber \\
&&  
= -\sqrt{\frac{\pi}{2}} b_{\beta}^{\dagger i}(\epsilon)d_{j}^{\dagger\beta}(\epsilon)\left|0\right\rangle _{\epsilon\rightarrow 0.} 
\eea

As the currents are traceless, we have to subtract the trace part for $i=j$. 
The latter can be neglected for large $N_c$. 

\bigskip

{\bf Comments:}

1. 
The adjoint vacuum can also have flavor quantum numbers. In fact, when using 
different flavor indices on the fermion and anti-fermion [and not summing], we 
can get a vacuum that is also adjoint in flavor. This will result in 
a ``currentball'' being an adjoint flavor multiplet as well.

2.
For the case of $N_f=N_c$, the state constructed from the bosonic adjoint 
vacuum is degenerate with the one constructed from the fermionic adjoint 
vacuum.

\bigskip

Denote
\beq
Z^a \equiv -\sqrt{\frac{2}{\pi}} (J^a)^{'} (0)
\eeq

The ``one current'' state we have in mind is  
\beq 
|k\rangle = J^{b}(-k) Z^b |0\rangle
\eeq 
This state is obviously a global color singlet, but in our Light Cone gauge   
$A_-=0$ it is also a local color singlet, as the appropriate line integral   
vanishes.  

Now  
\beq
\sqrt{\frac{\pi}{2}}\left[J^{a}(p),Z^{b}\right]=\frac{1}{2}N_{f}\delta^{ab}\delta(p)-if^{abc}  
(J^{c})^{'}(p) 
\eeq

From which, for $p>0$  
\beq
J^{a}(p)Z^{b}|0\rangle=0 
\eeq
Hence the state $Z^{b}|0\rangle$ is annihilated by all the annihilation  
currents, and so it is indeed a colored vacuum.  
  
From 
\beq
\left[P^{+},J^{b}(-k)\right]=kJ^{b}(-k)  
\eeq
we get that our state $|k\rangle$ is indeed of momentum $k$. 

\section{Evaluation of $M^2$}
\bea  
\lefteqn{\left[\int_{0}^{\infty}dp\phi(p)J^{a}(-p)J^{a}(p),J^{b}(-k)\right] = 
} \nonumber \\ 
&&  
\frac{1}{2} N_{f} \frac{1}{k} J^{b}(-k) + \nonumber \\
&& i f^{abc} \int_{0}^{k}{dp}\phi(p)J^{a}(-p)J^{c}(p-k) + \nonumber \\
&& if^{abc} \int_{k}^{\infty}dp\left(\phi(p)-\phi(p-k)\right) \nonumber \\
&&
J^{a}(-p)J^{c}(p-k)
\eea

In $P^{-}$ [and in $P^{+}$] we ignore contributions from  
zero - mode states, that is, we cut the integrals at $\epsilon$,  
and then take the limit. Actually, the zero modes give zero when 
acting on singlet states.   
  
\beq
P^{-}J^{b}(-k)Z^{b}|0\rangle =   
 \left[P^{-},J^{b}(-k)\right]Z^{b}|0\rangle 
\eeq 
as the Hamiltonian annihilates the color vacuum as well.  
  
Using the commutator of the Hamiltonian with a current, which we evaluated   
before, we get  
\bea
\lefteqn{
\frac{\pi}{e^2} P^{-} J^{b}(-k) Z^{b}|0\rangle =  \frac{1}{2} N_{f} \frac{1}{k}
 J^{b}(-k)Z^{b}|0\rangle +} \nonumber \\
&&
if^{abc} \int_{0}^{k}dp\phi(p)J^{a}(-p)J^{c}(p-k)Z^{b}|0\rangle 
\eea

\bea
\lefteqn{M^{2}J^{b}(-k)Z^{b}|0\rangle = 2P^{-}P^{+}J^{b}(-k)Z^{b}|0\rangle}
\nonumber \\
&& = 2k P^{-}J^{b}(-k)Z^{b}|0\rangle \nonumber \\
&& = (\frac{e^{2}N_{f}}{\pi})J^{b}(-k)Z^{b}|0\rangle+ 
(\frac{2e^{2}}{\pi}k)if^{abc}
\nonumber \\
&&
\int_{0}^{k}
\phi(p)J^{a}(-p)J^{c}(p-k)Z^{b}
|0\rangle 
\eea
So, in the large $N_{f}$ limit, the state $J^{b}(-k)Z^{b}|0\rangle$  
is an eigenstate, with eigenvalue $\frac{e^{2}N_{f}}{\pi}$.  
  
To see the exact dependence of the two terms, to be denoted I and II,
 in the equation above  
(the one and two current states), on $N_{f}$ and $N_{c}$, we should  
normalize them. For I,
\bea
\lefteqn{\left\langle 0\right|Z^{a}J^{a}(k)J^{b}(-k)Z^{b}\left|0\right\rangle 
=} \nonumber \\
&& 
\frac{1}{2}N_{f}k\delta(0)\left\langle 0\right|Z^{b}Z^{b}\left|0\right\rangle
   +  N_{c}\left\langle 0\right|Z^{b}Z^{b}\left|0\right\rangle 
\eea
The second term in the last line can be neglected compared to the first, 
as it   
is a constant compared with $\delta(0)$, the space volume divided by   
$2 \pi$.  
\beq
\langle0|Z^{b}Z^{b}|0\rangle = (N_c^2 - 1) \langle0|Z^{1}Z^{1}|0\rangle 
\eeq 
and the factor $k \delta(0)$ is the normalization of a plane wave of momentum 
k. 

The normalized state, for $N_c >> 1$,  
\beq  
\frac{1}{N_{c}\sqrt{\frac{1}{2}N_{f}}}J^{b}(-k)Z^{b}|0\rangle  
\eeq  
relative to $\langle0|Z^{1}Z^{1}|0\rangle$.  
  
The normalization of the second term II is more complicated.  
A lengthy but straightforward calculation gives  
\bea
\lefteqn{\left\| \left(if^{def} k 
\int_{0}^{k}dq\Phi(q)J^{d}(-q)J^{f}(q-k)\right)Z^{e}\left|0\right\rangle
\right\|^{2}= } \nonumber \\
&&
N_{c}\left(N_{c}^{2}-1\right)\left(\frac{1}{2}N_{f}\right)^{2} 
k \delta(0)\left\langle 0\left|Z^{1}Z^{1}\right|0\right\rangle \nonumber \\
&& \times
(Ints)  
\eea

\bea
\lefteqn{ (Ints)  = } \nonumber \\
&&
 k \int_{0}^{k}dpp(k-p)\Phi(p)\left(\Phi(p)-\Phi(k-p)\right) \nonumber \\
&& -k \frac{N_c}{N_f}\int_{0}^{k}dp\Phi(p)\int_{0}^{k-p}dq q
\Phi(q)
\eea
We have written only the terms proportional to $\delta(0)$. 

Useful formulae for the evaluation, involving sums of  
products of structure functions of $SU(N)$, are given in the Appendix. 
 
The various momentum integrals (including the ones for the non-dominant 
terms) are divergent for $\epsilon\rightarrow0$, thus they should 
be regulated. We leave this problem for now, and assume henceforth 
that they are regulated and finite. For simplicity the integrals  
(including the factor k) appearing 
in the two dominant terms will be notated $R_{1}$ and $- R_{2}$ in 
the following expressions. Note that we have $\frac{1}{\epsilon^2}$ and
$\frac{1}{\epsilon}$
divergences and also $\ln(\frac{k^2}{\epsilon^2})$.  
It seems that these are cancelled in $R_2$. 
 
 Define now the normalized states 
\beq 
|S_1\rangle = C_1 \left(J^{b}(-k)Z^{b}\left|0\right\rangle \right) 
\eeq
 
\beq
|S_2\rangle = i C_2 k f^{abc}\int_{0}^{k}dp\Phi(p)J^{a}(-p)J^{c}(p-k)Z^{b} 
\left|0\right\rangle 
\eeq
where 
\beq
C_1= \frac{1}{N_c\sqrt{\frac{1}{2}N_f}} \qquad  
C_2=\frac{ \frac {2}{N_f\sqrt{N_c^3}}}{\sqrt{R_1 + R_2\frac{N_c}{N_f}}} 
\eeq
 
The mass eigenvalue equation
\bea
\lefteqn{
M^2 |S_1\rangle = 
\frac{e^2}{\pi}N_{f} |S_1\rangle} \nonumber \\
&&
+ \frac{e^2N_c}{\pi} \sqrt{2} 
\sqrt{R_1\frac{N_f}{N_c} + R_2} |S_2\rangle
\eea
In the large flavor limit, our state $|S_1\rangle$ is  
an eigenstate with mass   
\beq
M= \sqrt {e^2 N_f \over {\pi}}.   
\eeq 
In the large color limit, however, we actually get   
that the   
second term dominates by a factor of $N_c$. Moreover, while the first term 
goes to zero in the large $N_c$ limit, due to the factor of $e^2$, the second 
term  
survives.

\section{Fermionic Structure}
 
Consider
\beq
J^i_j (-k)  b^{\dagger j}_\beta(\epsilon) d^{\dagger \beta} _i ( \epsilon)
\left|0\right\rangle  
\eeq

with
\bea
\lefteqn{J^i_j (-k) = \frac{1}{2\pi}\int_0^\infty dp}
\nonumber \\
&&
[ b^{\dagger i}_\beta(p+k) b^{\beta} _j (p) +  
\theta(k-p) b^{\dagger i}_\beta(k-p) d^{\dagger \beta} _j (p)  
\nonumber \\
&&
- d^{\dagger \beta}_j(p+k) d^{i} _\beta (p)] 
\eea
 The 4-quarks part has a coefficient which is independent of $N_c$. 
 As for the 2-quark part, it involves an anti-commutator of creation with 
annihilation, yielding a state which is a combination of  
$$
{b^{\dagger j}_\beta(k) d^{\dagger \beta} _j (\epsilon)},
$$
and
$$
b^{\dagger j}_\beta(\epsilon) d^{\dagger \beta}_j (k) 
$$
with a coefficient that is proportional to $N_c$. 

Thus for large $N_c$, we have a quark-antiquark combination of momenta 
$( k,0)$ and   $(0, k)$. 
  
As 't Hooft found all meson states for large $N_c$, and each has a well 
defined momentum distribution, it is clear that our state is not a mass 
eigenstate  
for large $N_c$.  This is of course part of our explicit calculation in the 
previous section.  
 
\section{Discussion}
Present work:

1. 
Investigation of  two dimensional massless multi flavor $QCD$ in its 
``currentized'' form. 

2. Useful tool  
to solve  those $QCD_2$ systems.  

3.
Spectrum of states that are constructed by applying a single 
current creation operator on the adjoint vacuum. 

4.
The later was shown to be given in terms of the derivative with respect 
to $k$, at $k = 0$, of the current acting on the singlet 
vacuum. 

5.
In general, and in particular also in the large $N_c$, these states are not 
eigenstates of $M^2$. However, 
in the large $N_f$ limit they are eigenstates.  

\bigskip

Previously, Ref \cite{arfs}, 
the spectrum of  two current states on the singlet vacuum was derived.
 It was also shown there that  
in the large $N_f$ limit  there is a continuum of states with mass above 
$2\times e\sqrt{\frac {N_f}{\pi}}$. 
This indicates that there is a non-interacting  ``currentball'' meson of 
mass $ e\sqrt{\frac {N_f}{\pi}}$. 

Now, in the same large $N_f$ limit, we have indeed found an eigenstate of 
$M^2$ 
with exactly  
this mass.
The state we have found is a color singlet. In fact it is easy to see that, in 
the   large $N_f$ limit, there
are   $N_c^2-1$ colored eigensates of $M^2$ with the same mass. 

Interpretation:  
For large $N_f$, $QCD_2$ is transfered into a set of $N_c^2-1$ Abelian 
systems. Now in  $QED_2$  it is well known that  the Schwinger mechanism  
yields a massive state of mass $\frac{e}{\sqrt{\pi}}$. 
This Schwinger state is often considered as a bound state of an 
electron-positron.  
For large $N_f$, the $M^2$ eigenstates are therefore just the Schwinger 
states appearing in a multiplicity of $N_c^2-1$. 
 
\bigskip

Open questions: 

1. 
The computation of the spectrum of the states requires a regularization 
prescription 
that we have only partially found.  

2.
Diagonalize the ``currentball'' states created  by applying any number of  
current creation operators on  the various vacua.

3.  
The question of what have we learned from the currentization procedure and 
from the  
two dimensional spectrum of states, about four dimensional $QCD$. 

4.
In particular a challenging question is to investigate the 
possibility of a Schwinger like mechanism also in four dimensions.  

\section{Appendix - Summation Identities}
 
The generators $T^a$ of $SU(N)$, in the adjoint representation, are related to 
the structure constants $f^{abc}$ as  
$$ (T^a)_{bc} = -i f^{abc}$$ 
Thus  
\beq
f^{abc} f^{abd} = Tr (T^c T^d) = N \delta^{cd}
\eeq
and  

\bea
\lefteqn{ f^{abc} f^{a'bc'} f^{aa'd} f^{cc'd} 
= Tr (T^b T^d T^b T^d) } \nonumber \\
&& =i f^{bde} Tr (T^e T^b T^d) + Tr (T^b T^b T^d T^d)  \nonumber \\
&& = {1\over 2} N^2 (N^2 -1)
\eea
where we used 
\beq
\sum_a T^a T^a = N I_{adj}
\eeq
with $I_{adj}$ the unit matrix in the adjoint representation.

\end{document}